\documentclass{emulateapj}

\newcommand{\mstar}{$M_{\star}$\,}
\newcommand{\mb}{$M_{B}$\,}
\newcommand{\msun}{$M_{\sun}$\,}
\newcommand{\rbreak}{$R_{Br}$\,}
\newcommand{\rscale}{$R_{s}$\,}

\newcommand{\hone}{$h_{1}$\,}

\newcommand{\magarcsq}{mag/$arcsec^{2}$\,}
\newcommand{\bb}{$B_{435}$\,}
\newcommand{\vv}{$V_{606}$\,}
\newcommand{\ii}{$i_{775}$\,}
\newcommand{\zz}{$z_{850}$\,}
\newcommand{\R}{$R$\,}

\slugcomment{No comments}

\shorttitle{Color Profiles of Disk Galaxies}
\shortauthors{Azzollini, Trujillo \& Beckman}

\begin{document}


\title{Color Profiles of Disk Galaxies since z$\sim$1: Probing Outer Disk Formation Scenarios}


\author{R. Azzollini, I. Trujillo and J. E. Beckman\altaffilmark{1}}
\affil{Instituto de Astrof\'isica de Canarias (IAC), C/V\'ia L\'actea s/n, 38205 La Laguna, S/C de Tenerife, Spain}
\email{ruyman@iac.es, trujillo@iac.es, jeb@iac.es}


\altaffiltext{1}{Consejo Superior de Investigaciones Cient\'ificas, Spain, and IAC}


\begin{abstract}
We present deep color profiles for a sample of 415 disk galaxies within the redshift range 0.1$<$$z$$\leq$1.1, and contained in HST-ACS imaging of the GOODS-South field \citep{Giavalisco04}. For each galaxy, passband combinations are chosen to obtain, at each redshift, the best possible aproximation to the rest-frame $u-g$ color. We find that objects which show a truncation in their stellar disk (Type II objects) usually show a minimum in their color profile at the break, or very near to it, with a maximum to minimum amplitude in color of $\lesssim$0.2 \magarcsq, a feature which is persistent through the explored range of redshifts (i.e. in the last $\sim$8 Gyr). This color structure is in qualitative agreement with recent model expectations \citep{Roskar08} where the break of the surface brightness profiles is the result of the interplay between a radial star formation cut-off and a redistribution of stellar mass by secular processes.
\end{abstract}


\section{Introduction}\label{sec1}
Systematic studies of the fainter outer parts of stellar disks in galaxies show that there are three distinct modes of behaviour. All disk galaxies show radial exponential decay in their surface brightness profiles \citep[][]{Patterson40,deVaucouleurs59}, though in some cases this exponential may change its slope (i.e. scale length) at a defined galactocentric radius. First there are the pure Type I profiles, with no change of slope in their surface brightness profile out to 5 and even 10 scale lengths \citep{Bland-Hawthorn05, PT06, Erwin08}; second, the Type II profiles, with a shallower inner exponential followed by a steeper outer exponential \citep{Freeman70, Erwin05, PT06, Erwin08}. A subset of this type are the so called stellar disk ``truncations'' \citep{Kruit79}, where the change in slope takes place in the outermost parts of the disks \citep[see also][]{Pohlen04}; and finally, the Type III, seen in  \citet{Courteau96} and in \citet{MatthewsGallagher97}, but first systematically measured by \citet{Erwin05}, and observed in many galaxies by \citet{PT06} and by \citet{Erwin08}. Type III profiles have a steeper inner exponential and a shallower outer profile and have been termed ``antitruncations''.

In understanding why and how these differences amongst profiles arise, the appliance of stellar population analysis techniques throughout stellar disks may prove very useful. This kind of analysis, for example, sheds light on when stars formed in different parts of the disk of galaxies, so giving hints on the stellar mass build-up process. When we can apply these tools on objects at different redshifts, we may see how changes in the spatial distribution of stellar mass in the disks actually take place. Color gradients have been extensively used in the literature as a way to extract information on both ages and  metallicities through the galaxy radial profiles in the Local Universe \citep[e.g.][]{MacArthur04}. However, an interesting point these previous works missed was the connection between the color distribution and the break phenomenology. The reason for this is that although truncations in edge-on galaxies are known since quite long, a systematic exploration showing the different galaxy profile types (i.e. Type I,  II and III) in low inclination galaxies is relatively new, as accounted for above. At higher redshift (0.5$<z<$3.5), color distribution has also been studied (e.g. Moth \& Elston 2002 in the HDF-N or  Tamm \& Tenjes 2006 in the HDF-S) showing  for a general mixed population (ellipticals, spirals and irregulars) a small to a constant color gradient. But again, these previous works were not focused at all on the break phenomenology. In a significant step towards tackling this problem, we present here the analysis of color profiles for 415 disk galaxies taken from the GOODS-South field with profile Types I, II, and III in the redshift range 0.1$<z\leq$1.1. The observed HST/ACS bands were selected to best approximate rest-frame $u-g$, for maximum uniformity in the interpretation. The results put useful constraints on models of disk formation in general, more evidently for profile truncations.

Throughout, we assume a flat $\Lambda$-dominated cosmology ($\Omega_{M}$ = 0.30, $\Omega_{\Lambda}$=0.70, and $H_{0}$=70 km $s^{-1}$ $Mpc^{-1}$). All magnitudes are in the AB system.
\section{DATA, SAMPLE SELECTION AND COLOR PROFILES}\label{sec2}
In \citet{ATB08} (ATB08 hereinafter) we presented a detailed and extensive analysis of surface brightness profiles for a sample of 505 late-type galaxies with redshifts 0.1$<$$z$$\leq$1.1, within the GOODS-South field. The final set of 435 galaxies in ATB08 that were suitable for analysis constitutes our parent sample. These galaxies were selected from the catalog published in \citet{Barden05}, which relies on the \emph{Galaxy Evolution from Morphologies and SEDs} imaging survey \citep[GEMS;][]{Rix04}. \citet{Barden05} provided morphological analysis of that sample of galaxies by fitting S\'ersic $r^{1/n}$ \citep{Sersic68} profiles to their surface brightness. Also, photometric redshift estimates (with errors $\delta$$z$/(1+$z$)$\sim$0.02) and absolute $B$-band magnitudes of the objects were available thanks to the photometric catalog from COMBO-17 \citep[\emph{Classifying Objects by Medium-Band Observation in 17 filters};][]{Wolf01, Wolf03}. ATB08 selected those objects from the Barden et al. sample which fell within the following ranges of parameters: S\'ersic index $n$$\leq$2.5 to isolate disk-dominated galaxies \citep{Ravindranath04}; axial ratio $q$ $>$ 0.5 to select objects with inclination $<$ 60$\arcdeg$; and $M_{B}$ $<$ -18.5 magnitudes, as in \citet{TP05}. Moreover, only objects with z$<$1.1 were selected in order to maintain the analysis in the optical rest-frame bands. The observational basis for this work and ATB08 is imaging data from HST-ACS observations of the GOODS-South field \citep{Giavalisco04}\footnote{\url{http://www.stsci.edu/ftp/science/goods/}}. The data set consist of images in the $F435W$, $F606W$, $F775W$ and $F850LP$ HST pass-bands, hereafter referred to as \bb, \vv, \ii and \zz. As in ATB08, we divide the redshift range covered by the objects in three redshift bins: ``low'': 0.1$<$$z$$\leq$0.5; ``mid'': 0.5$<$$z$$\leq$0.8 and ``high'': 0.8$<$$z$$\leq$1.1. All objects are classified as being of Type I, II or III, following ATB08. This classification was obtained by inspection of the GOODS band image which best approximates to rest-frame $B$-band in each redshift bin.

In order to make the best comparisons of the color profiles in different redshift ranges we use the images in the filter sets which give the closest approximations possible to rest-frame $u-g$, which are, for 0.1$<$$z$$\sim$0.5  \bb-\vv, for 0.5$<$$z$$\sim$0.8 \vv-\ii, and for 0.8$<$$z$$\sim$1.1 \vv-\zz.

The parent sample of 435 galaxies was further reduced by eliminating probable S0's and those Type II profiles where the change in slope clearly occurs in the inner disk (i.e. in the region inner to the spiral arms). We were also forced to remove 10 galaxies (4 of Type I and 6 of Type II) in the nearest redshift bin for which there is no \bb image, since the field for this filter is slightly smaller than for the others. The final sample for this study has 415 galaxies. In Table \ref{tblClass} we present the distribution of objects amongst the three redshift bins (``low'', ``mid'' and ``high''; in rows), and profile Types (I, II and III; in columns), according to the described selection process.

Details of how the radial brightness profiles were derived can be found in ATB08, and the color profiles were subsequently produced as the difference between the surface brightness profiles in the two appropriate bands. The surface brightness profiles were radially scaled using a scale radius \rscale, whose definition depends on the profile Type, to give a uniform radial coordinate \R. Objects of Types II or III have \rscale equal to the break radius \rbreak. For Type I objects \rscale is defined as a factor $f_{h}$ times the disk scale length h. ATB08 measure \rbreak/\hone for Type II galaxies of intermediate redshift as 1.4, where \hone is the inner scale length, while for nearby galaxies \citet{PT06} measure a median value of 2.4 for this ratio. These values led us to choose $f_{h}$=2 so that the radial scaling is comparable for Types I and II. To give a sense of typical aparent/proper sizes of the galaxies under study, for the subset of Type II objects, the median values of \rbreak are 1.33''/5.25 kpc, 0.75''/5.17 kpc and 0.64''/5.07 kpc within the redshift bins termed as ``low'', ``mid'' and ``high'' respectively. We have also checked the effect of the different ACS PSF wing properties on our color gradients, finding basically non influence on our results.\\

\begin{figure}
\epsscale{1}
\plottwo{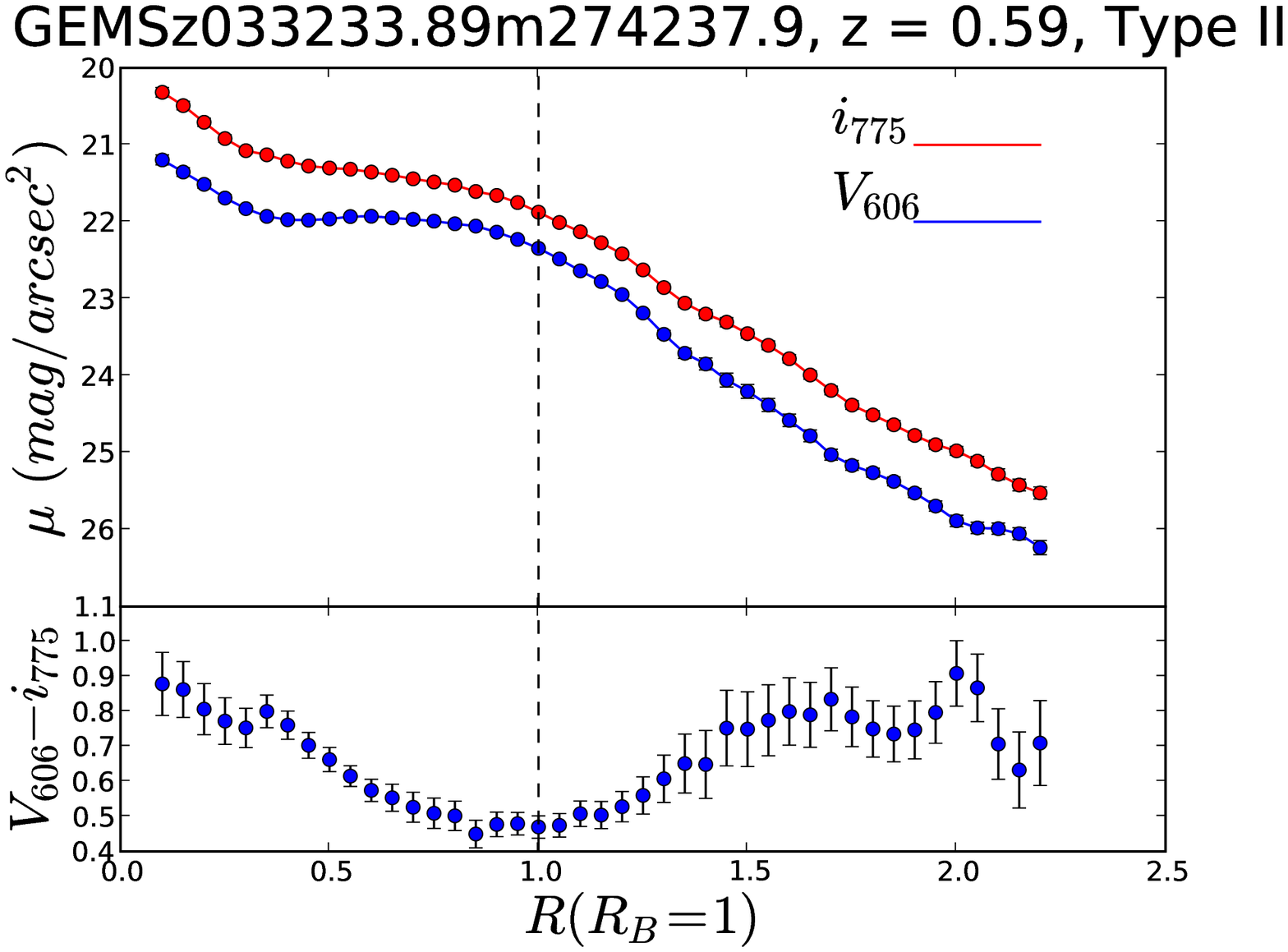}{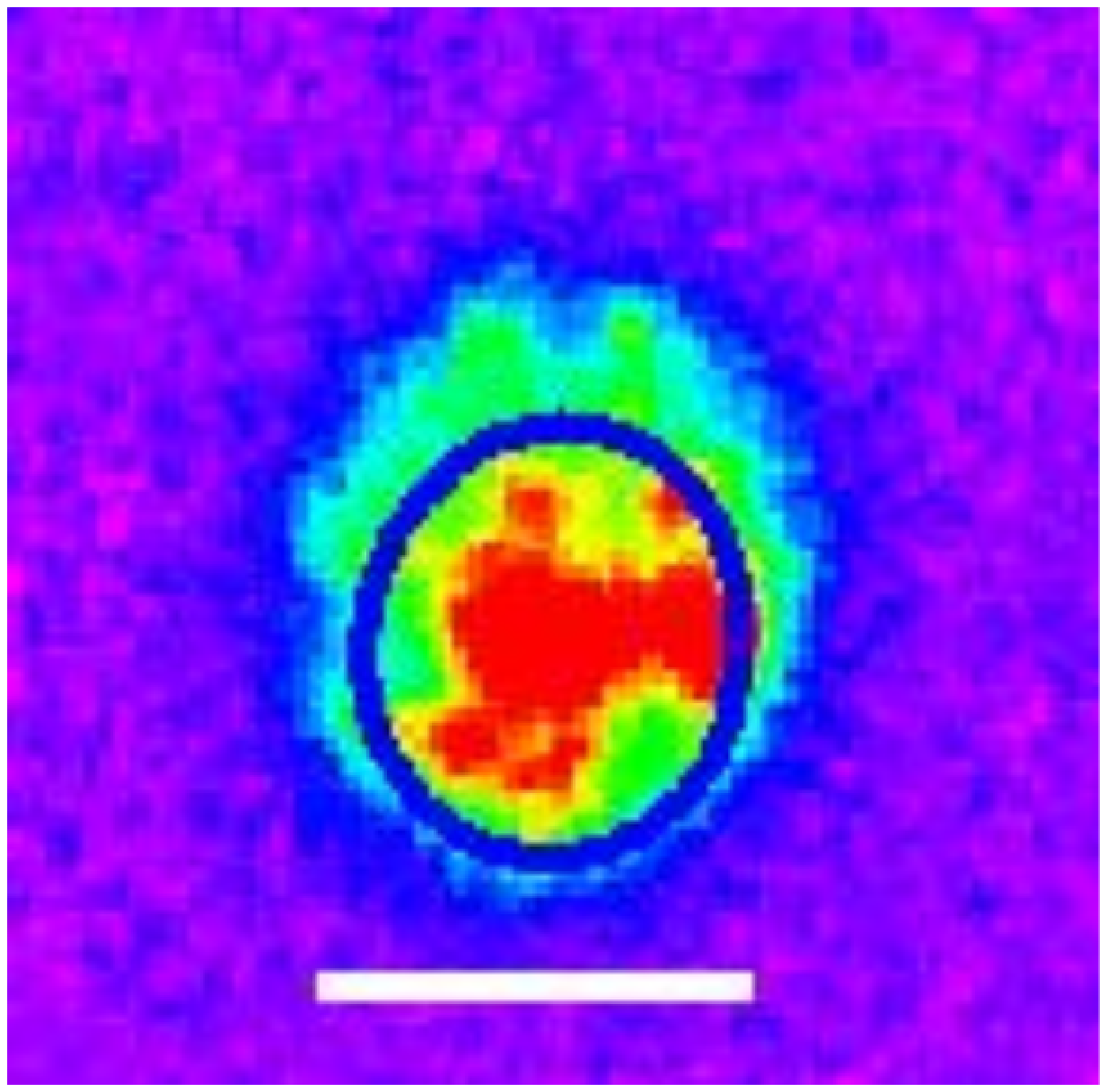}
\caption{Ilustrative example of the kind of analysis performed on the galaxies of the sample under study, for a Type II galaxy at z=0.59. In the upper left panel, surface brightness profiles in the bands \ii and \vv. In the same lower left panel the \ii - \vv color profile, showing a clear minimum in color around \R=1 (corresponding to \rbreak). On the right panel we show the \ii image of the galaxy from GOODS-South-HST/ACS. The ellipse marks the position of the break in the surface brightness profile. The horizontal line is 1 arcsec long at the given scale.\label{figExample}}
\end{figure}

To illustrate our procedure, in Fig. \ref{figExample} we present an example color profile for one of the objects in the sample. It is a Type II galaxy (ATB08) at $z$=0.59. There is a broad but clear minimum (bluer) color at the position of the break (\R$\sim$1).
\section{RESULTS AND DISCUSSION}\label{sec3}
\begin{figure}
\epsscale{1}
\plotone{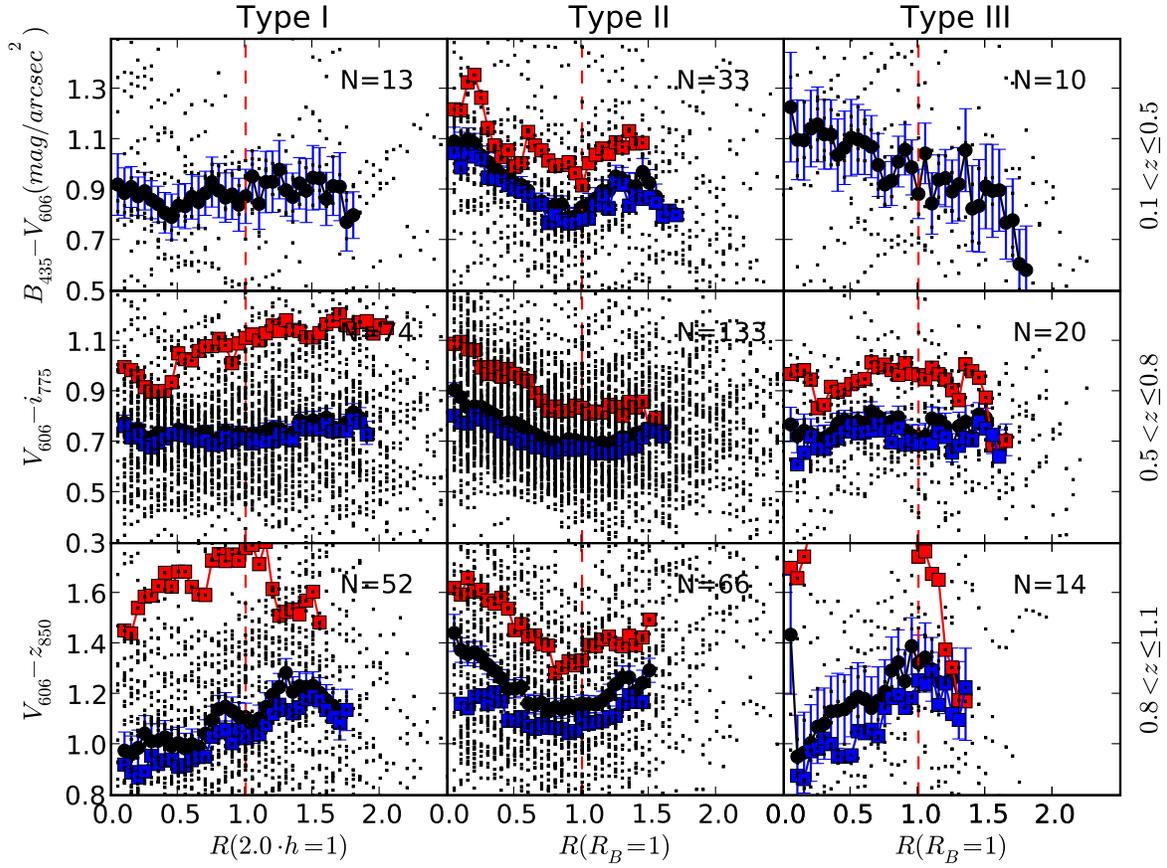}
\caption{Color profiles of the 415 galaxies under study. The sample is divided in sub-samples according to surface brightness profile Type (I, II or III, in columns, from left to right) and redshift range (``low'', ``mid'' or ``high'', in rows, from up to down). The colors (\bb-\vv, \vv-\ii, \vv-\zz) are chosen as the best proxies to the rest-frame u-g color in each redshift bin. The radii are scaled the scale radius, \rscale, whose definition depends on profile type: \rscale=2$\cdot$h for Types I, where h is the scale length of the disk, and it is equal to the break radius, \rscale=\rbreak, for Types II and III. Small points are individual color profiles. Big black dots are the median color profiles for each sub-sample, and the error bars give the error in those estimations. The error in the median color profile $\delta$$C$ is defined as $\delta$$C$=$\sigma_{C}$ / $N^{1/2}$, where $\sigma_{C}$ is the standard deviation in the distribution of color at the given radius, and N is the number of galaxies taken into account, which is also shown in the panels. The red squares give the median color profile for objects with stellar mass \mstar$>$$10^{10}$\msun, while the blue squares are the analogous for objects with \mstar$\leq$$10^{10}$\msun.\label{figColor}}
\end{figure}

Using our color profiles for the 415 sample galaxies we computed the median color profiles for sub-samples of Type I, II, and III, divided into bins of ``low'', ``mid'' and ``high'' redshift, all displayed in Fig. \ref{figColor}. We show these profiles for all nine subclasses, with their error bars, and also, coded in color, median profiles for galaxies in higher and lower ranges of stellar mass \mstar (see caption for details).\\

In Fig. \ref{figColor} we see that in all three redshift bins the Type II profiles show color minima at the break radius. The differences $\Delta$$C$ in median color index $C$ between \R=0 and \R=1 are, for the three bins, in increasing redshift 0.14$\pm$0.05 \magarcsq, 0.17$\pm$0.02 \magarcsq and 0.20$\pm$0.05 \magarcsq respectively, a marginal increase with $z$. The reported errors $\delta\Delta$$C$ are $\sqrt{2}$ times the median of the error in the median color index median($\delta$$C$)$_{R}$ through radii (The $\sqrt{2}$ factor comes from the propagation of the error in the median color index when obtaining $\Delta$$C$ as a difference in color). The error in the median color index $\delta$$C$ is obtained as the standard deviation of the color profiles at each radius, divided by the square root of the number of profiles. These differences in color index $\Delta$$C$ are 2.6, 7.3 and 3.9 times higher than the errors $\delta\Delta$$C$, in the respective bins, and so the (blue) minima are clearly significant. The minimum values are found at $R$=0.9, 1.1, and 0.85, respectively, i.e. close to, but not exactly at \R=1 in each case. The minima are, as we see in Fig. \ref{figColor}, specific to Type II profiles.

For Type I galaxies the median color profiles show no feature close to \R=1 (\rscale=2$\cdot$$h$). They are nearly flat in the two nearer redshift bins, and more bumpy, rising with \R, (redder with increasing \R) but with no clear feature, in the furthest bin. Type III profiles in the two nearer bins are linear with \R, falling (going bluer) with \R in the nearest bin, and flat in the second bin. Interestingly in the furthest bin the profile has a maximum (reddest) at \R=1, apparently opposite to Type II behaviour. Nonetheless, this must be taken with caution, as this sub-sample does not have a big population (number of galaxies $N$=14).

There is a large scatter between individual profiles for all Types and bins. Among the possible causes are the quite wide range of $z$ within each bin, causing a scatter in the effective band-passes, the range of inclinations, causing different amounts of dust reddening, and intrinsic scatter due to different evolutionary stages for the galaxies in a bin. We could check among these by plotting the color at \R = 1 against the parameters: a) 1/$q$, where $q$ is the axis ratio of the ellipse best fitting the disk; b) redshift, c) \mb, (absolute mag in B) and d) the stellar mass \mstar of the galaxy ; \citep[c) and d) are from][]{ Barden05}. The strongest dependences of $C$(\R=1) are on the stellar mass, \mstar, for most profile Types and at all redshifts (the only exception being for Type I galaxies in the ``low'' redshift bin, a sample with $N$=13). We have also tested for the relations between $C$(\R=1) and average B-band luminosity surface density ($L_{\sun}/kpc^{2}$) and average stellar mass surface density (\mstar/$kpc^{2}$), both within the radius corresponding to \R=1. These relations are not stronger, though in the case of stellar mass surface density comparable, to that found with stellar mass. Thus, the color scatter is due mainly to differences in stellar mass, and not to the inclination range nor to omitting $k$-corrections for the range of z within each bin.

This is illustrated in Fig. \ref{figColor} where we have plotted separately, where meaningful,(not in the low redshift bin for Types I and III where the samples were too small) median color profiles for objects with 
\mstar$\leq$$10^{10}$\msun (blue squares) and \mstar$>$$10^{10}$\msun (red points), superposed on the plots of the full samples (black dots). The full profiles lie close to the ``low mass'' profiles, while the high mass profiles lie well above the other two. Our sample is dominated by the lower mass objects, and there are significant differences between low and high mass galaxies. However the segregated plots for the Type II´s all reproduce the color minima shown by the full plots. We note that in the highest redshift bin the Type III profiles of high mass objects are very much redder than those of the less massive ones, possibly indicating contamination of this sub-sample with early type galaxies.

The most interesting feature of our results is the minimum in the color profile at the break for Type II galaxies, with similar amplitude over the full redshift range and for the stellar mass range explored. This seems to confirm that truncation is a phenomenon related to the stellar populations of the galaxies, and not a purely structural or geometrical effect.

Proposed models to explain the occurrence of truncations in stellar disks can be grouped in two branches: a) models related to angular momentum conservation in the protogalactic cloud \citep{Kruit87} or angular momentum cut-off in cooling gas \citep{Bosch01}; and b) models which appeal to thresholds in star formation \citep{Kennicutt89} or star forming properties of distinct ISM phases \citep{ElmegreenParravano94,Schaye04,ElmegreenHunter06}. More recently, N-body simulations in which both star formation processes and secular dynamics are included, have been devised in an attempt to better fit observational results \citep[e.g.][]{Debattista06}. In this same line, \citet{Roskar08} presented a model in which the breaks are the result of the interplay between a radial star formation cut-off and a redistribution of stellar mass by secular processes. In this model, stars are created inside the break (threshold) radius and then move outwards due to angular momentum exchange in bars and/or spiral arms. Our results fit qualitatively their prediction that the youngest stellar population should be found at the break radius, and older (redder) stars must be located beyond that radius. It is not easy to understand how ``angular momentum'' or ``star formation threshold''/``ISM phases'' models alone could explain our results. Thus they pose a difficult challenge for these models. However, it will be necessary to check whether the \citet{Roskar08} models are able to reproduce quantitatively the results shown here.

There are also other possible causes for the color minima which cannot be ruled out, but at this stage are less satisfactory. Radially falling dust thickness out to the break plus older stars outside it could give this result for the color, but does not explain the break itself. More subtly, a change in metallicity gradient at the break could explain the color profile, but again this would need incorporating into a predictive model to yield a break.

The color profiles for Types I and III are less easy to interpret. Type I's show flat profiles in the lower redshift bins, and a rising profile in the furthest bin. In the latter case this could imply a greater star formation rate near the centers. Type III's show a falling profile in the nearest bin, a flat profile in the intermediate bin, and a rising profile with a downturn at the break in the highest redshift bin, this latter being the inverse of the behavior of the Type II's. We do not feel that it is worth hazarding even qualitative interpretations for Type I's and Type III's, certainly not within the scope of this letter and especially given the small or moderate numbers involved. But the results, when strengthened statistically, will surely lead to useful constraints on disk evolution models, preferably when compared with similar results from more local samples (as in e.g. Bakos et al., in preparation).
\acknowledgments
We are grateful to Marco Barden for kindly providing us with the GEMS morphological analysis catalog. We thank the COMBO-17 collaboration (especially Christian Wolf) for the public provision of a unique database upon which this study is based. We must acknowledge the GOODS team for providing such a valuable database from which we could extract our results. We also thank the anonymous referee for fruitful comments from which the final result of this work has significatively benefited. This work is based on observations made with the NASA/ESA \emph{Hubble Space Telescope}, which is operated by the Association of Universities for Research in Astronomy, Inc., under NASA contract NAS5-26555. Partial support has been provided by projects AYA2004-08251-C02-01 and AYA2007-67625-C02-01 of the Spanish Ministry of Education and Science, and P3/86 of the Instituto de Astrof\'isica de Canarias.
{\it Facilities:} \facility{HST (ACS)}

\clearpage

\begin{table}
\begin{center}
\caption{Sample distribution by profile type and redshift ranges}\label{tblClass}
\begin{tabular}{r|rrrrrr}
\tableline\tableline
                    &  I   &  II   & III   &  Total  &  $Total_{ATB08}$ & Discarded\\
\tableline
$0.1 < z \leq 0.5$  &  13  &   33  &   10  &    56   &    67	   &	  11 (10)  \\
$0.5 < z \leq 0.8$  &  74  &  133  &   20  &   227   &   234	   &	  7     \\
$0.8 < z \leq 1.1$  &  52  &   66  &   14  &   132   &   134	   &	  2     \\
\tableline
Sum                 & 139  &  232  &   44  &   415   &   435	   &	  20 (10)  \\
\tableline
\end{tabular}
\tablecomments{The total number of objects in each redshift bin is given in column 4. In column 5 we give the number of objects within each redshift bin of the parent sample (from ATB08), and in column 6 how many were discarded to produce the ``child'' sample. Also in the column 6, in parenthesis, and when it applies, is given the number of objects that were discarded because there is no data in one of the bands to produce the desired color.}
\end{center}
\end{table}

%% the file, uncomment the \plotone, \plottwo, and
%% If you need a layout that cannot be achieved with \plotone or
%% \plottwo, you can invoke the graphicx package directly with the
%% \includegraphics command or use \plotfiddle. For more information,
%% please see the tutorial on "Using Electronic Art with AASTeX" in the
%% documentation section at the AASTeX Web site,
%% http://www.journals.uchicago.edu/AAS/AASTeX.
%%
%% The examples below also include sample markup for submission of
%% supplemental electronic materials. As always, be sure to check
%% the instructions to authors for the journal you are submitting to
%% for specific submissions guidelines as they vary from

%% This example uses \plotone to include an EPS file scaled to
%% 80% of its natural size with \epsscale. Its caption
%% has been written to indicate that additional figure parts will be
%% available in the electronic journal.

\end{document}